\documentclass[12pt]{article}\usepackage{etex}
\usepackage[textheight=9in, textwidth=6.5in, letterpaper]{geometry}

\usepackage[utf8]{inputenc}
\usepackage{epsfig}
\usepackage{multirow}

\usepackage{amsfonts}
\usepackage{amscd}
\usepackage{latexsym}
\usepackage{bbold}
\usepackage{amsmath,amssymb}
\usepackage{verbatim}
\usepackage{tcolorbox}
\newtcolorbox{mymathbox}[1][]{colback=white, sharp corners, #1}
\usepackage{setspace}
\usepackage{hhline}
\usepackage{physics}
\usepackage{color}
\usepackage{cite}
\usepackage{hyperref}
\usepackage{graphicx}
\usepackage{tikz}
\usepackage{tikz-3dplot}
\usetikzlibrary{calc}
\usetikzlibrary{positioning}
\usepackage{float}
\usepackage{subfig}
\usepackage[T1]{fontenc}
\usepackage{amsmath, amssymb, amsthm, fullpage}
\usepackage{doc}
\usepackage{ifthen, color, fancyvrb}
\usepackage{nextpage}

\usepackage{wrapfig}
\usepackage{parskip}
\usepackage{authblk}
\usepackage[makeroom]{cancel}

\numberwithin{equation}{section}

\def\be{\begin{equation}}
\def\ee{\end{equation}}

\def\p{\partial}

\def\be{\begin{equation}}
\def\ee{\end{equation}}
\def\bea{\begin{eqnarray}}
\def\eea{\end{eqnarray}}
\def\<{\langle }
\def\>{\rangle}

\begin{document}
\begin{titlepage}
\unitlength = 1mm

\vskip 15cm

\begin{center}
\vspace{9cm}
{ \LARGE {\textsc{Soft Gravitons  in the BFSS Matrix Model}}}

\vspace{1cm}
Noah Miller, Andrew Strominger, Adam Tropper and Tianli Wang

\vspace{0.4cm}

{\it  Center for the Fundamental Laws of Nature, Harvard University,\\
Cambridge, MA 02138, USA}\\

\begin{abstract}
BFSS proposed that asymptotically flat M-theory is dual to a large $N$ limit of the matrix quantum mechanics describing $N$ nonrelativistic D0-branes.  Recent insights  on the soft symmetries of any quantum theory of gravity in asymptotically flat space are applied to the BFSS matrix model.  It is shown that soft gravitons are realized by submatrices whose rank is held fixed in the large $N$ M-theory limit, rather than the usual linear scaling with $N$ for hard gravitons. The soft expansion is  identified with the large $N$ expansion and the soft theorem becomes  a  universal formula for the quantum mechanical scattering of such submatrix excitations.   This formula is shown to be the Ward identity of large type IIA $U(1)_{RR}$ asymptotic  gauge symmetry in the matrix model, whose asymptotic boundaries are at future and past timelike infinity. 

\end{abstract}

\vspace{1.0cm}

\end{center}

\end{titlepage}

\pagestyle{empty}
\pagestyle{plain}

\def\vx{{\vec x}}
\def\p{\partial}
\def\po{$\cal P_O$}

\pagenumbering{arabic}

\tableofcontents

\section{Introduction}

The holographic principle has a concrete and well-understood realization in anti-deSitter space (AdS) \cite{Maldacena:1997re}. One hopes that the principle extends in some form to (nearly) flat spacetimes like the one we inhabit. The basic fact that the ratio of the boundary to bulk volume goes to a constant at large radius in AdS
and to zero in flat space suggests that flat space holography may differ qualitatively from its AdS counterpart. But exactly how is an outstanding open question. 

Two seemingly  different approaches to flat space holography are the Banks-Fischler-Shenker-Susskind (BFSS) matrix model \cite{deWit:1988wri,Banks_1997, Susskind:1997cw, Seiberg_1997,Sen:1997we, Polchinski:1999br, Taylor_2001, bigatti1999review, Ydri:2017ncg} and 
celestial holography \cite{deBoer:2003vf,He:2015zea,Pasterski:2016qvg,Strominger:2017zoo,Raclariu:2021zjz,Pasterski_2021}. BFSS is a top-down construction equating the momentum-$N$ sector of  discrete lightcone quantized (DLCQ) M-theory with a quantum mechanics  of $N\times N$ hermitian matrices representing open strings stretching between $N$ D0-branes. Celestial holography is a bottom-up approach applicable to any quantum theory of gravity in flat space, including M-theory, in which the proposed dual field theory lives on the celestial sphere at null infinity.   Since the two approaches  are applicable  to the same theory it is natural to explore their connection.  

The starting point  of celestial holography (as well as AdS holography) is that both sides of a dual pair must have the same symmetries. Given the bulk description, soft theorems provide an efficient route to finding these symmetries 
\cite{Strominger:2017zoo}. So the first question we ask in this paper  is `Is the soft graviton theorem realized in BFSS?' We answer this by showing that soft gravitons are matrix subblocks whose rank is held fixed (rather than scaling with $N$ like the hard gravitons) in the large-$N$ limit which recovers the full uncompactified M-theory.\footnote{This is reminiscent of the large-$N$ limit of QCD, where baryons have masses of order $N$ and mesons  of order $1$. It would be interesting to see how far this analogy can be pushed.} The soft limit is then nontrivially identified with the M-theory limit. It would be illuminating to derive the soft theorem directly from the matrix model, and would provide a novel test of the latter. 

Soft theorems are in general expected to be Ward identities of symmetries. Hence one asks if this expectation holds for the soft theorem in the matrix model. Using the known expression for  the BFSS matrix model in a background $U(1)_{RR}$
gauge field \cite{Taylor:1999gq} we show that the soft theorem is the Ward identity of `large' $U(1)_{RR}$ gauge transformations \cite{Strominger:2013lka} which do not die off at past or future timelike infinity.\footnote{This is in accord with the fact that 11D supertranslations with non-zero momentum on the M-theory circle KK reduce to $U(1)_{RR}$ gauge transformations \cite{Marotta:2019cip, Ferko:2021bym}.}

We hope the answers to these basic questions provide a jumping off point for relating these two approaches to flat holography.  Many further questions remain unanswered. 

We will begin by reviewing the BFSS matrix model in Section \ref{sec: BFSS matrix model}. In Section \ref{sec: soft theorem}, we leverage the soft graviton theorem in M-theory into an analogous one  in the matrix model dual. We demonstrate that the soft expansion in the matrix model is a $1/N$ expansion.  In Section \ref{sec: RR Gauge Field}, we discuss the interplay between soft theorems and supertranslation symmetry in M-theory arguing that the analog of supertranslation symmetry in the matrix model is a large gauge symmetry of the RR 1-form.

\section{Matrix Model Review}
\label{sec: BFSS matrix model}
In this section we briefly review the relevant features of the BFSS matrix theory \cite{deWit:1988wri,Banks_1997, Susskind:1997cw, Seiberg_1997,Sen:1997we, Polchinski:1999br, Taylor_2001, bigatti1999review, Ydri:2017ncg}, which 
conjectures that the compactification of M-theory on a lightlike circle $X^- \sim X^- + 2 \pi R$ with momentum $P^+ = N/R$ is dual to the low-energy dynamics of $N$ D0-branes in 10 dimensions or, equivalently, a certain supersymmetric quantum mechanical theory  of $N\times N$ Hermitian matrices. 
Readers familiar with BFSS may safely skip this section. 

\subsection{BFSS Duality}

Compactification of 11-dimensional M-theory on a spacelike circle gives type IIA string theory \cite{Townsend:1995kk, Witten:1995ex}. 
The massless degrees of freedom in M-theory  are the  11-dimensional supergraviton multiplet. 
The D0-branes in type IIA string theory are identified as the KK-modes of this  supergraviton multiplet.  The number of units $N$ of  momentum around the  circle corresponds to the number of D0-branes. The BFSS matrix model concerns a lightlike compactification of M-theory which can be defined as an infinitely large boost of a spacelike one \cite{Seiberg_1997}.

Let us define the lightcone coordinates $X^\pm,$ and lightcone momenta $P^{\pm}$ by
\begin{equation}
    X^\pm = \frac{1}{\sqrt{2}}(X^0 \pm X^{10}), \hspace{30pt} P^{\pm} = \frac{1}{\sqrt{2}}(P^0 \pm P^{10}).
\end{equation}
Lightcone quantization is performed on surfaces of constant $X^+$ which plays the role of time, with $P^-$ the Hamiltonian. The lightlike compactification of M-theory is
\begin{equation}
    (X^+,X^-) \sim (X^+, X^- + 2\pi R).
\end{equation}
$P^+$ is quantized according to  
\begin{equation}\label{qu}
  P^+ = N/R.
\end{equation}

BFSS argued that the sector of M-theory with total momentum $P^+ = N/R$ can be described by a rescaled version of the Hamiltonian encoding the low-energy dynamics of $N$ $D0$-branes in type IIA string theory \cite{deWit:1988wri,Banks_1997, Susskind:1997cw}
\begin{equation}
    H = \frac{R}{2}\text{Tr}\Bigg[P^I P^I - \frac{1}{2(2\pi l_p^3)^2}[X^I,X^J][X^I,X^J] - \frac{1}{2\pi l_p^3}\Psi^T \Gamma^I[X^I,\Psi]\bigg]
    \label{eqn: BFSS Hamiltonian}
\end{equation}
subject to the constraint on physical states
\begin{equation}
  f_{ABC} (X^I_B P^I_C - \frac{i}{2} \Psi^\alpha_{B} \Psi^\alpha_{C})|\psi_{\text{phys}}\rangle = 0
\end{equation}
which forces states to be invariant under $U(N)$ transformations.
Here $X^I$ are $N \times N$ Hermitian matrices with the index $I = 1,...,9$ running over the directions transverse to the lightlike compactification. $P^{I}$ are their conjugate momenta. $\Psi^\alpha$ is an $N \times N$ Hermitian matrix-valued spinor of $Spin(9)$ with $\alpha = 1,...,16$ and gamma matrices $\Gamma^I_{\alpha \beta}$. One can decompose these matrices as 
\begin{equation}
    X^I = X^{I}_A T^A, \hspace{30pt} P^I = P^{I}_A T^A, \hspace{30pt} \Psi^\alpha = \Psi^\alpha_{A} T^A
\end{equation} where $T^A$ are generators of the Lie algebra of $U(N)$  in the adjoint representation normalized so that  $\text{Tr}(T^AT^B) = \delta^{AB}$. 


We now review some basic properties of this theory \cite{Danielsson:1996uw, Kabat:1996cu, Bachas:1995kx, Sethi_1998, moore2000d, yi1997witten}. The bosonic potential $V \sim \Tr([X^I, X^J]^2)$ is classically at a minimum $V = 0$ when $[X^I,X^J]=0$ for all $I$ and $J$, which implies all matrices can be simultaneously diagonalized. The $N$ eigenvalues are  then positions of the $N$ D0-branes. For example, $N$ non-interacting D0-branes travelling along trajectories $x_{i}^I(t)$ with $i = 1,...,N$ are described by the diagonal matrices
\begin{equation}\label{diag_matrix}
    X^I(t) = \begin{pmatrix} x^I_1(t) & 0 & \ldots & 0 \\0  & x^I_2(t) & \ldots & 0 \\ \vdots & \vdots & \ddots & \vdots \\ 0 & 0 & \ldots & x^I_N(t)\end{pmatrix}.
\end{equation}
The off-diagonal elements are open strings stretching between the D0-branes. 

A clump of $m<N$ coincident D0-branes corresponds to an $m \times m$  sub-block of this matrix for which all of the eigenvalues $x_i^I$ in the sub-block are equal. Quantum mechanically, these clumps are marginally bound states with complicated wave functions. They are dual to the higher KK-momentum supergraviton modes in the M-theory picture. 
Widely separated clumps are noninteracting because the  strings stretched between them are very massive and forced into their ground states. 

\subsection{Scattering }

Consider a set of $n$ gravitons in M-theory with individual momenta $k_{j}^+ = N_j/R,$ with $j = 1,...,n$ and total momentum $P^+_{\text{tot}} = (N_1+\cdots + N_n)/R = N/R.$ Each graviton is dual to a marginally bound clump labeled by the number of D0-branes $N_j$, the transverse momentum $k^I_j$, and the polarization information of the 11D supergraviton multiplet $\epsilon_j$ which in the D0-brane description is encoded by the trace `center of mass' fermions, with the explicit map given in \cite{plefka1998quantum,Plefka_1998,Plefka:1998in,Plefka:1997hm}. The dictionary \cite{Banks_1997, Susskind:1997cw, Seiberg_1997, Becker:2006dvp} between clumps of D0-branes and a collection of M-theory gravitons is: 
\begin{table}[H]
    \centering
    \begin{tabular}{|c c c|}
    \hline
         \textbf{M-Theory} & ~ & \textbf{BFSS}  \\
    \hline
         $k_{j}^+$ & $\Longleftrightarrow$ & $N_j/R$ \\
         $k_j^I$ & $\Longleftrightarrow$ & $k_j^I$ \\
         $k_{j}^-$ & $\Longleftrightarrow$ & $R(k_j^I)^2/2N_j$ \\
         \hline
    \end{tabular}
    \caption{Dictionary between momenta of gravitons in M-theory and momenta of D0-brane clumps in the BFSS matrix model. The final relation is determined using the mass-shell condition for M-theory gravitons $0 = -2k_j^+k_j^- + k_j^I k_j^I$}
    \label{tab: Kinematics Dictionary}
\end{table}

Widely separated multi-graviton `scattering states'  in M-theory with quantum numbers $k_1^\mu,\epsilon_1,...,k_n^\mu,\epsilon_n$ correspond to widely-separated multi-clump states in the matrix model with D0-brane quantum numbers $N_1,k_1^I,\epsilon_1,...,N_n,k_n^I,\epsilon_n$. 
In order to formulate the scattering problem in BFSS, one should consider initial and final scattering states corresponding to widely separated wavepacket clumps of D0-branes, evolving past into future using the BFSS Hamiltonian \eqref{eqn: BFSS Hamiltonian} \cite{Becker:1997wh, Becker:1997xw}. The BFSS duality conjecture states that this scattering amplitude matches with the one that one would compute in (lightlike compactified) M-theory, namely
\begin{equation}
    \mathcal{A}_{\text{M}}(k_1^\mu,...,k_n^\mu) = \mathcal{A}_{\text{BFSS}}(N_1,k_1^I,...,N_n,k_n^I).
    \label{eqn: M-theory and BFSS Amplitude Relation}
\end{equation}
\subsection{The M-theory Limit}
The scattering amplitudes of uncompactified 11-dimensional $M$-theory are  obtained by taking the radius large with external momenta held fixed 
\be \label{mone}R\to \infty, ~~~~~~~k^\mu_j ~~{\rm fixed}.\ee
From the expression for the momenta this is easily seen to be equivalent to 
\be \label{mtwo}N\sim N_j \sim R \to \infty,  ~~~~~~~k^I_j ~~{\rm fixed},\ee
and hence is a variety of large-$N$ limit. In this limit, the discretuum of allowed values of external momentum approach a continuum and  any scattering proccess can be studied. 

\begin{figure}
    \centering
    \includegraphics[width = \textwidth]{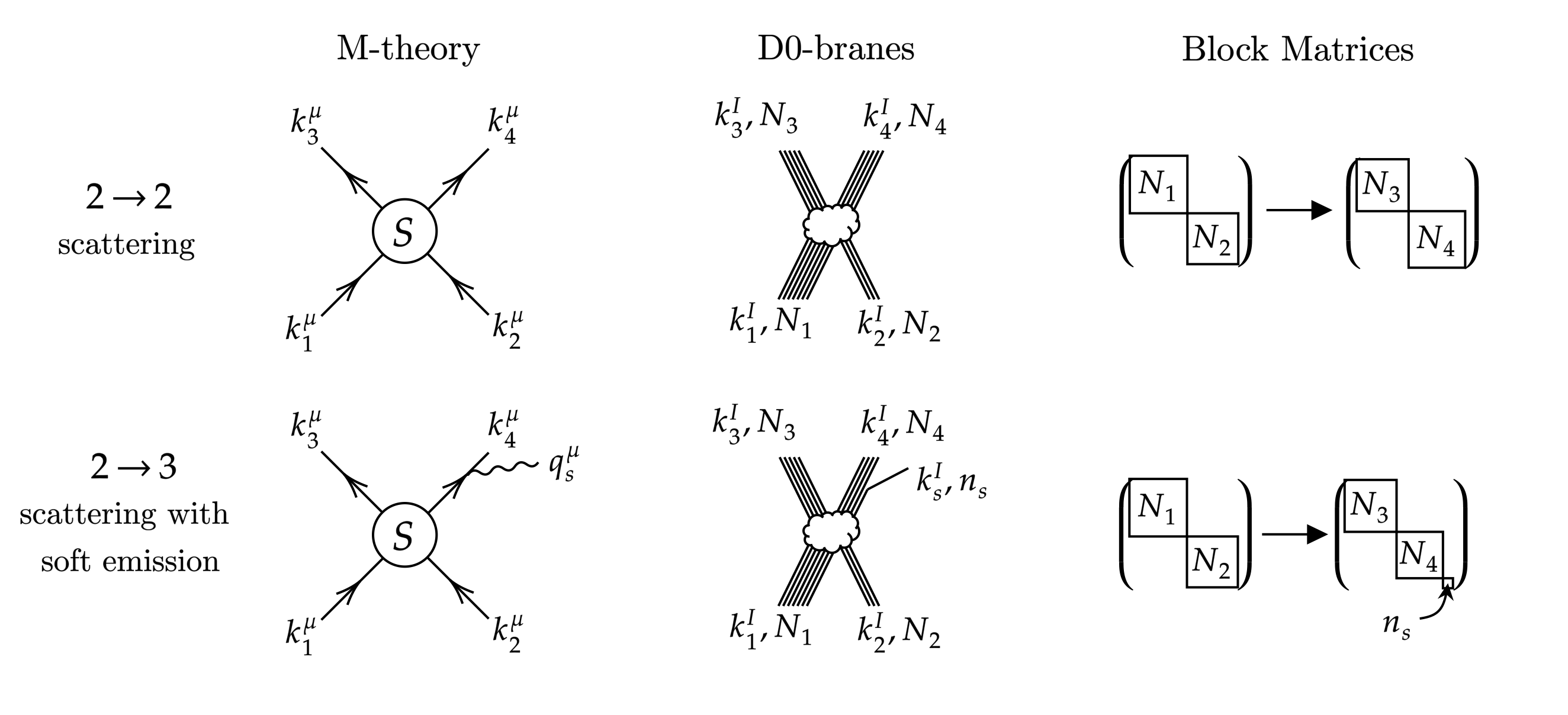}
    \caption{A schematic depiction of equivalent scattering processes viewed through dual lenses of M-theory, D0-brane interactions, and block diagonal matrices in the BFSS model.}
    \label{fig:ScatteringProcesses}
\end{figure}

\section{Soft Graviton Theorem in the Matrix Model}
\label{sec: soft theorem}

In this section we show that the soft theorem is realized within  BFSS duality and, moreover, that the soft limit is the same as the M-theory limit with external momenta/D0-charges suitably scaled. 
\subsection{The Soft Theorem in M-Theory}

Weinberg's soft graviton theorem \cite{Weinberg:1965nx}  applies to essentially any gravitational theory in an asymptotically flat spacetime.\footnote{Of course Weinberg considered only four dimensions, but the extension to higher dimensions is straightforward  \cite{Kapec_2017,He_2019_1,He_2019,Kapec_2022,Marotta:2019cip}

 }   In particular  it must hold in 11-dimensional M-theory which contains gravitons as part of the low energy effective action.

Consider a generic scattering amplitude $\mathcal{A}_{\text{M}}(k_1^\mu,...,k_n^\mu)$ involving external particles with future-directed momenta $k^\mu_j$. These external particles may be ingoing or outgoing  gravitons or some other particles.  Momenta are parameterized by a vector $v^I \in \mathbb{R}^d$ and a scale $\omega \in \mathbb{R}_{>0}$ according to 
\begin{equation}
    k^\mu_j = \omega_j \hat{k}^\mu_j = \frac{\omega_j}{2}(1 + v_j^2, 2 v_j^I, 1 - v_j^2), \hspace{30pt} q^\mu_s = \omega_s \hat{q}^\mu_s = \frac{\omega_s}{2}(1 + v_s^2, 2 v_s^I, 1 - v_s^2) . 
    \label{eqn: graviton parametrization}
\end{equation}
where we denote the momentum of the soft graviton by $q^\mu_s$. The soft graviton theorem states:
\begin{equation}
    \mathcal{A}_\text{M}(q^\mu_s,\epsilon_s^{\mu \nu};k_1^\mu,...,k_n^\mu) = \Bigg[\frac{\kappa}{2} \epsilon^s_{\mu \nu} \sum_{j =1}^n \eta_j \frac{k^\mu_j k^\nu_j}{q_s \cdot k_j} + \mathcal{O}\left(\Big(\frac{\omega_s}{\omega_j}\Big)^0\right) ~ \bigg] \mathcal{A}_\text{M}(k_1^\mu,...,k_n^\mu)
    \label{eqn: soft theorem in gravity}
\end{equation}
where $\eta_j = +1 ~ (-1)$ if the $j^{\rm th}$ particle is outgoing (incoming), $\epsilon_s^{\mu \nu}$ is the polarization tensor of the soft graviton, and $\kappa = \sqrt{32\pi G_N}$ \cite{Weinberg:1965nx}.

In the soft limit,  the ratios $\omega_s/\omega_j \rightarrow 0$. The coefficient of the leading soft divergence $(\omega_s/\omega_j)^{-1}$ is universal. This leading soft term has corrections which are a power series in $(\omega_s/\omega_j)$.

\subsection{Soft Limit in the Matrix Model}
\label{sec: Soft Limit}

In this subsection  we show  the soft limit in the BFSS  matrix model is  the M-theory limit with  hard gravitons represented by subblocks whose size grows like $N$ and soft gravitons by subblocks of fixed finite size. 

Both soft and hard gravitons are parameterized as in Equation (\ref{eqn: graviton parametrization}) with $q_s^+ = \sqrt{2}\omega_s$ and $k_j^+ = \sqrt{2}\omega_j$. This expression implies
\begin{equation}
    \frac{n_s}{R} = \sqrt{2} \omega_s ~ ~ , ~ ~ q_s^I = \omega_s v_s^I \sim \omega_s \hspace{20pt}\text{and}\hspace{20pt} \frac{N_j}{R} = \sqrt{2} \omega_j ~ ~ , ~ ~ k^I_j = \omega_j v_j^I \sim \omega_j
    \label{eqn: soft parametrization}
\end{equation}
where we have used the dictionary provided in Table \ref{tab: Kinematics Dictionary} with $N_j$, the block sizes, corresponding to the hard M-theory gravitons, and $n_s$ to the soft ones. Thus, the momentum $k_{j}^+$ for a particular graviton dictates the size of the corresponding block in the matrix model. In the matrix model, the soft limit then reads
\begin{equation}
    \textbf{Soft Limit:} ~ ~~~ ~ ~ ~ \frac{n_s}{N_j} = \frac{\omega_s}{\omega_j} \rightarrow 0.
    \label{eqn: block ratio}
\end{equation}
Scattering amplitudes with a soft external particle in the BFSS matrix model, thus, correspond to situations where a block of size $n_s$ is dwarfed by the other blocks of size $N_j$.  This happens automatically in the M-theory limit \eqref{mtwo} as long as we keep $n_s$ fixed! Hence the soft limit is the same as the M-theory limit, but with a new type of external state  constructed from a finite number $n_s$ of D0-branes. 

In the M-theory limit, the difference between scattering a graviton with $(N_j,k^I_j)$ versus  $(N_j-1,k^I_j)$  with one fewer  D0-branes vanishes. This might have led to the naive conclusion that submatrices with sizes or order one don't matter and that the scattering of a single D0-brane $(1, k^I_s)$ vanishes altogether. This is not the case because of the soft pole. Note also the leading term in the scattering amplitude for a bound state with a fixed finite number $n_s$ of D0s differs only by the multiplicative factor $1/n_s$.


We illustrate a $2 \rightarrow 3$ scattering process with soft emission diagrammatically from the M-theory perspective, the D0-brane perspective, and the block diagonal matrix perspective explicitly in Figure \ref{fig:ScatteringProcesses}.


We now write the leading soft graviton theorem of M-theory  \eqref{eqn: soft theorem in gravity} in terms of BFSS variables. If we define a convenient basis for graviton polarization tensors
\begin{equation}
    \epsilon^{\mu \nu}_{IJ}(v) \equiv \frac{1}{2}\big(\epsilon^\mu_I \epsilon^\nu_J + \epsilon^\nu_I \epsilon^\mu_J\big) - \frac{1}{d}\delta_{IJ}\epsilon^\mu_K \epsilon^{K\nu} \hspace{20pt} \text{with} \hspace{20pt} \epsilon^\mu_J(v) \equiv \partial_J \hat{q}^\mu_s = (v_J,\delta^I_J,-v_J)
\end{equation}
and write the soft graviton polarization as
\begin{equation}
    \epsilon_s^{\mu \nu} = e^{IJ} \epsilon_{IJ}^{\mu \nu},
\end{equation}
then after some algebra and using the dictionary \ref{tab: Kinematics Dictionary}, the soft theorem becomes \footnote{There is a small technical subtlety in Equation \eqref{eqn: leading soft theorem 1}. The BFSS matrix model describes M-theory in a sector with momentum $P^+ = N/R$, so all amplitudes must be manifestly momentum conserving in $P^+$ and cannot be off-shell in $P^+$. Equivalently, the number of D0-branes $N$ is always conserved. Therefore, we cannot simply append a small block of size $n_s$ to the matrix, but we must shrink the size of the other blocks slightly. Assuming that the amplitudes are analytic in $N_j$ (in the large $N$ limit, this follows from the analyticity of the M-theory S-matrix) one may perform a first order Taylor expansion to see that the expression will only be corrected at subleading terms.}
\begin{equation}
        \mathcal{A}_{\text{BFSS}}(n_s,q^I_s, \epsilon_s,\text{out; in}) = \bigg[-2 \kappa \sum_{j=1}^n \eta_j \frac{N_j}{n_s} \frac{e_{IJ}(v_s-v_j)^I (v_s-v_j)^J}{(v_s-v_j)^2} + \cdots ~ \bigg] \mathcal{A}_{\text{BFSS}}(\text{out; in}).
        \label{eqn: leading soft theorem 1}
\end{equation}
Finally, we define the inversion tensor in 9 spatial dimensions as\footnote{This is the same inversion tensor familiar from conformal field theory.}
\begin{equation}
    \mathcal{I}^{IJ}(v) = \delta^{IJ} - 2 \frac{v^I v^J}{v^2}.
    \label{eqn: polarization basis}
\end{equation}
In terms of this inversion tensor, the leading soft graviton theorem in the BFSS matrix model reads
\begin{equation}
    \mathcal{A}_{\text{BFSS}}(n_s,q^I_s,\epsilon_s,\text{out; in}) = \bigg[\kappa \sum_{j=1}^n \eta_j \frac{N_j}{n_s} e_{IJ} \mathcal{I}^{IJ}(v_s-v_j) + \cdots ~ \bigg] \mathcal{A}_{\text{BFSS}}(\text{out; in}).
    \label{eqn: leading soft theorem 2}
\end{equation}
Note  that the soft pole $\omega_j/\omega_s$ gets recast into the ratio of block sizes $N_j/n_s$, which diverges in the soft limit according to Equation (\ref{eqn: block ratio}).

Sub-leading corrections to this expression are given by an expansion in $n_s/N_j$. Because $n_s \sim \mathcal{O}(1)$ and $N_j \sim \mathcal{O}(N),$ we can identify the subleading terms in the soft expansion on the gravity side with a $1/N$ expansion on the gauge theory side.

It would be illuminating  to derive the soft theorem directly from the matrix model. It is not obvious to us even how the factor of $N_j/n_s$ would emerge. 

\section{Asymptotic Symmetries in the Matrix Model}
\label{sec: RR Gauge Field}

In this section, we use the soft graviton theorem in 11D to show that the insertion of a single  D0-brane  in a 10D BFSS scattering amplitude generates a large gauge transformation on the background RR 1-form gauge potential $C_\mu$ in the matrix model.\footnote{The result easily generates to finite bound clumps of D0-branes by dividing by $n_s$. } Since this is a quantum-mechanical model the relevant asymptotic regions are at $t=\pm \infty$. The RR 1-form is  of the form $C_\mu = \partial_\mu \theta_{e, v_s}$, for some particular gauge parameter $\theta_{e, v_s}$ given in \eqref{thta} depending on the polarization $e_{IJ}$ and velocity $v_s$ of the soft D0-brane. This is summarized in Equation \eqref{final_eq}, which is the main result of this section. This large $U(1)$ gauge symmetry arises in the KK reduction of the 11D supertranslation symmetry.\footnote{Symmetries associated to 10D supertranslations would have to come from modes independent of the $X^+$ circle and hence involve  $n_s=0$. It is not clear to us how to describe  these in  the matrix model.}

\subsection{Background RR Gauge Potentials}

The standard BFSS matrix model, with the Hamiltonian given by Equation \eqref{eqn: BFSS Hamiltonian}, describes a system of D0-branes living in world where all background fields are turned off.  The effect of coupling the D0-branes to external background fields can be incorporated by adding terms to the Lagrangian. In particular the interaction term coupling the D0-branes to the $U(1)_{RR}$ gauge field $C_\mu$ generalizes  the usual electromagnetic interaction $Q \int dt \; \dot x^\mu C_\mu(x)$ between a charge $Q$  particle and the gauge field, where $x^\mu$ is the worldline of the particle.

In the matrix model, the precise interaction term was found in \cite{Taylor:1999gq} to be \begin{equation}\label{wati_action}
   S_{RR}[C_\mu] = \int dt \sum_{n = 0}^\infty \frac{1}{n!} \big(\partial_{I_1} \cdots \partial_{I_n} C_\mu(t, \vec{0})\big)  I^{\mu (I_1 \cdots I_n) }
\end{equation}
where  $\mu = 0, \ldots, 9$ and  $x^\mu = (t, x^I) = (t, \vec{x})$. 
The `multipole moments' of the current $I^\mu$ are defined by
\begin{align} \label{eqn: current}
    I^{\mu (I_1 \cdots I_n)} 
    &= \Tr (\mathrm{Sym}(I^\mu, X^{I_1}, \cdots , X^{I_n} )) +I^{\mu (I_1 \cdots I_n)}_{\text{F}} 
\end{align}
where
\begin{equation}
    I^\mu = (\mathbb{1}/R,\dot X^I/R).
\end{equation}
Here, $\mathrm{Sym}$ is a symmetrized average over all orderings of the input matrices. $I^{\mu (I_1 \cdots I_n)}_{\text{F}}$ are terms involving at least two fermionic matrices, $\Psi$, which will not be relevant to this paper for reasons discussed in section \ref{sec: asymptotic symmetry}. If one takes the matrices $X^I$ to be diagonal, as in Equation \eqref{diag_matrix}, then the action reduces to the electromagnetic form, as expected. 
\subsection{Large $U(1)_{RR}$ Gauge Transformations}
\label{sec: asymptotic symmetry}
If the RR 1-form is pure gauge, then the interaction term \eqref{wati_action} becomes a total derivative. Plugging $C_\mu = \partial_\mu \theta$ into Equation \eqref{wati_action}, one can show that\footnote{To demonstrate this, one must use that the multipole moments satisfy the conservation law
$\partial_t I^{0 (I_1 \cdots I_n)}=I^{I_1 (I_2 \cdots I_n)} + \cdots + I^{I_n (I_1 \cdots I_{n-1})}$.}
\begin{equation}
    \begin{split}
    S_{RR}[\partial_\mu \theta] &= \frac{1}{R} \int dt ~ \partial_t \left[ \sum_{n = 0}^\infty \frac{1}{n!} (\partial_{I_1} \cdots \partial_{I_n} \theta(t, \vec{0} )) I^{0 (I_1 \cdots I_n)} \right] \\
    &= \frac{1}{R} \sum_{n=0}^\infty \frac{1}{n!}(\partial_{I_1} \cdots \partial_{I_n} \theta(t,\vec{0})) I^{0(I_1 \cdots I_n)}\bigg|_{t = -\infty}^{t = +\infty}.
    \label{eqn: total derivative}
    \end{split}
\end{equation}
Therefore, a pure gauge background field  affects amplitudes by a position-dependent phase acting on the initial and final states.

The asymptotic symmetry group of gauge theories is typically defined as the set of `large' gauge transformations which satisfy some set of boundary conditions modulo  `small' gauge transformations which vanish at the boundary. For the remainder of this section, we will consider the case where $C_\mu = \partial_\mu \theta$ is pure gauge and given by such a large gauge transformation $\theta$ which is non-vanishing as $t \rightarrow \pm \infty.$ The boundary conditions which $\theta$ must satisfy near past and future timelike infinity \cite{He:2014cra,Campiglia:2015lxa,Kapec:2015ena,Campiglia:2015qka, Strominger:2017zoo} specify that as $t \rightarrow \pm\infty$, the gauge parameter $\theta(t,\vec{x})$ can only depend on the ratio $\vec{x}/t$ 
\begin{equation}
    \theta(t,\vec{x}) \xrightarrow{t \rightarrow \pm \infty} \theta(t, \vec{x}) = \theta(\vec{x}/t)
\end{equation}
implying that these large gauge transformations are parameterized by a single function on $\mathbb{R}^9$.\footnote{This is the limit relevant for nonrelativistic charged massive scattering states of the more general formula for large $U(1)$ gauge transformations. } Outside of the above specification, the gauge parameter is arbitrary.

Now we show that the boundary term \eqref{eqn: total derivative} reduces to a very simple expression on asymptotic scattering states. As functions of $X^I$, asymptotic  scattering  wavefunctions are  sharply peaked in momentum space and  non-trivially supported only  on matrices of the form
\begin{equation}
    X^I = \begin{pmatrix} x^I_1(t) & & \\ & \ddots & \\ & & x^I_N(t) \end{pmatrix} + \Delta X^I, \hspace{0.75 cm} \Delta X^I \sim \mathcal{O}(t^0)
    \label{eqn: asy states}
\end{equation}
where $\vec{x}_i(t) = \vec{v}_it + \vec{x}_{i,0}$ tracks the position of the $i^{th}$ D0-brane. Note that $\vec{x}_i(t) = \vec{x}_{j}(t)$ when the $i^{th}$ and $j^{th}$ D0-branes share a bound state. $\Delta X^I$ is a matrix whose values do not grow with time as $t \to \pm \infty$. The entries within the blocks on the diagonal of $\Delta X^I$ correspond to the degrees of freedom of the bound states modulo their center of mass motion. As such, these entries can take values on the order of the spatial size of these bound states. 

\begin{figure}
    \centering
    \begin{minipage}[b]{0.3\textwidth}
        \centering
        \includegraphics[width=\textwidth]{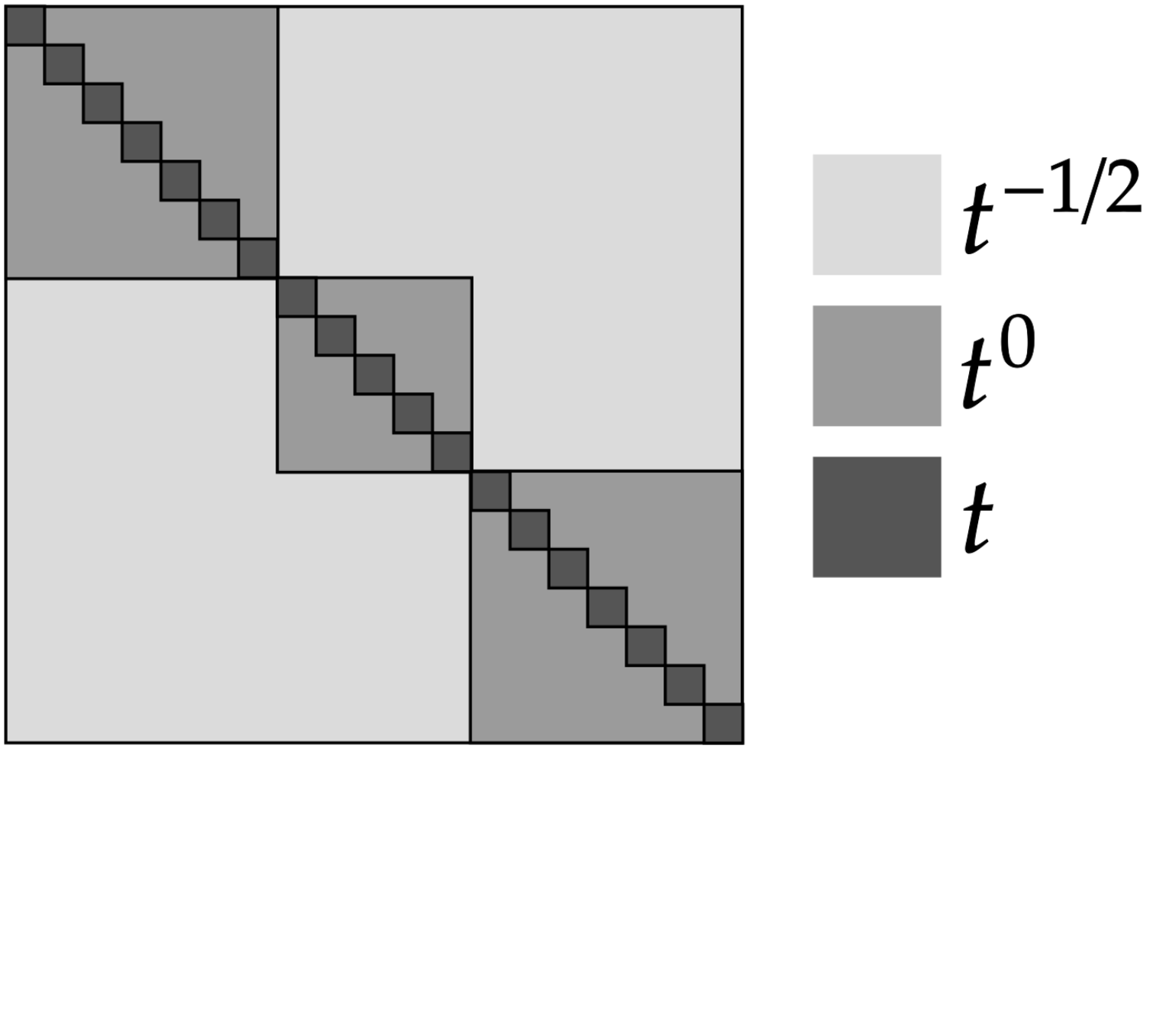}
        \caption*{(a)}
    \end{minipage}
    \begin{minipage}[b]{.65\textwidth}
        \centering
        \includegraphics[width=\textwidth]{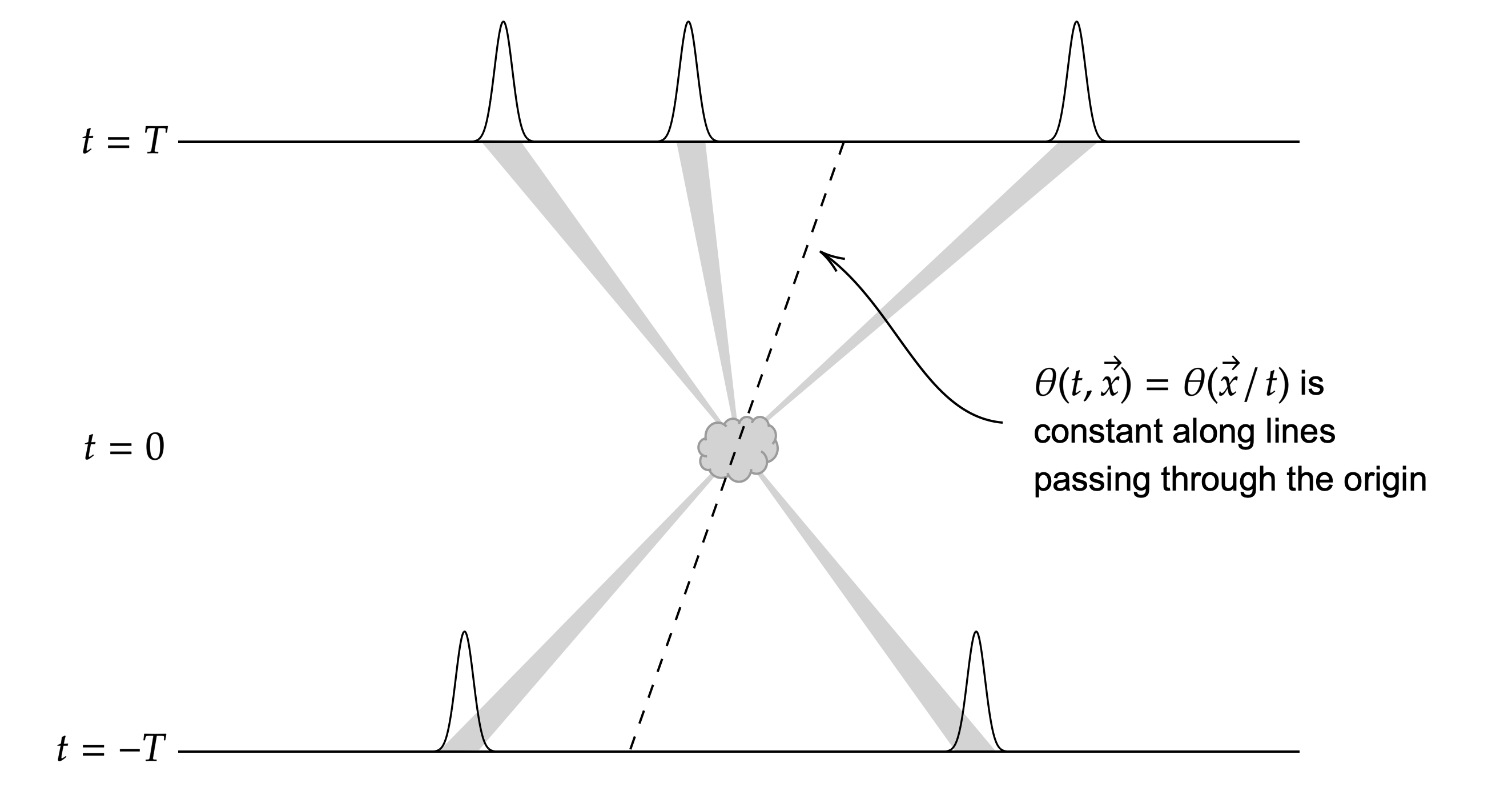}
        \caption*{(b)}
    \end{minipage}
    \caption{\label{wavepackets} (a) How different parts of the matrix $X^I$ scale with $t$ for scattering states. (b) The scattering of non-relativistic D0-brane bound state Gaussian wavepackets, which can be taken to have an arbitrarily small angular width. A gauge function $\theta(t,\vec{x})$ which only depends on the ratio $\vec{x}/t$ will take a well-defined value on these wave packets, depending only on their velocity.}
\end{figure}

The off block-diagonal components of $\Delta X^I$ describe strings stretched between distant D0-brane bound states. The mass of the string is proportional to its length, so these strings have mass scaling like $t$. When the string excitations become heavy, the wavefunction for these components gets frozen to the ground state of a quantum (super)harmonic oscillator with frequency $\omega \sim t$. The width of such a wavefunction shrinks as $\sim t^{-1/2}.$ This situation is summarized in Figure \ref{wavepackets}.

We may now insert the matrices describing the asymptotic states (Equation \eqref{eqn: asy states}) into the boundary term (Equation \eqref{eqn: total derivative}). We find that the only terms that survive at $t = \pm \infty$ are 
\footnote{In this footnote we demonstrate why the fermionic term $I_{\text{F}}^{0 (I_1 \ldots I_n)}$ in Equation \eqref{eqn: current} doesn't contribute to our analysis. First, we notice that the couplings $\partial_{I_1} \cdots \partial_{I_n} \theta(\vec{x}/t)$ contains $n$ spatial derivatives, each pulling down a factor of $1/t$. So, for the term $(\partial_{I_1} \cdots \partial_{I_n} \theta)I^{0(I_1 \cdots I_n)}$ to be non-vanishing at $t = \pm \infty$, $I^{0(I_1 \cdots I_n)} \sim t^n$ at asymptotic times. The bosonic moments, $I^{0(I_1 \cdots I_n)}_{\text{B}} = \text{Tr}(\text{Sym}(X^{I_1} \cdots X^{I_n})),$ scale as $t^n$, since every bosonic matrix has entries growing as $t$. In fact, \textit{only} the entries linear in $t$, which are $v^I t$, survive in this limit. Next, we use the fact that the BFSS action is invariant under $R \mapsto \lambda R, X \mapsto \lambda^{1/3} X, \Psi \mapsto \lambda^{1/2} \Psi, t \mapsto \lambda^{-1/3}t$ \cite{douglas1997d}. This invariance must persist when coupled to a background field if we map $\theta \mapsto \lambda \theta$, which implies that each multipole term $I^{\mu(I_1 \ldots I_n)}$ must have a constant scaling dimension in $\lambda$ for the action to be invariant. Therefore, if two $\Psi$'s are added, three $X$'s must be removed, making the whole term scale three lower powers of $t$ and vanish at the boundary.
}

\begin{align}
    S_{RR}[\partial_\mu \theta] & = \sum_{j = 1}^n \eta_j \frac{N_j}{R} \sum_{n=0}^\infty \frac{1}{n!} \big(\partial_{I_1}\cdots \partial_{I_n}\theta(t,0)\big) x_j^{I_1} \cdots x_{j}^{I_n} \bigg|_{t = -\infty}^{t = + \infty} = \sum_{j = 1}^n \eta_j \frac{N_j}{R} \theta(\vec{v}_j)
\end{align}
where we noticed that middle expression is just a Taylor expansion of $\theta(t,\vec{x})$ and used $\theta(t,\vec{x}_j) = \theta(\vec{x}_j/t)$ according to our earlier considerations.

In a quantum amplitude, this addition to the action becomes an overall phase. Therefore, placing BFSS in such a background gauge field modifies the amplitude via
\begin{equation}\label{eq_C_A}
        \eval{\mathcal{A}_{\text{BFSS}}(\text{out; in})}_{C_\mu = \partial_\mu \theta} = \exp( i \sum_{j = 1}^n \eta_j \frac{N_j}{R} \theta(\vec{v_j}) ) \eval{\mathcal{A}_{\text{BFSS}}(\text{out; in})}_{C_\mu = 0}.
\end{equation}
From the soft graviton theorem \eqref{eqn: leading soft theorem 2}, if we define the gauge parameter $\theta_{e, v_s}$ by
\begin{equation} \label{thta}
    \theta_{e, v_s}(\vec{x}/t) \equiv \frac{e_{IJ} (v_s - x/t)^I (v_s - x/t)^J}{|\vec{v}_s - \vec{x}/t|^2 }
\end{equation}
which depends on the velocity $\vec{v}_s$ and polarization structure $e_{IJ}$ of the soft D0-brane, then by combining \eqref{eq_C_A} and \eqref{eqn: leading soft theorem 1}, we see that
\begin{equation}\label{final_eq}
    \lim_{n_s/R \to 0} ~ \frac{n_s}{R} \eval{\mathcal{A}_{\text{BFSS}}(n_s, k^I_s, \epsilon_s,  \text{out; in})}_{C_\mu = 0} = - i \eval{\frac{d}{d \varepsilon}}_{\varepsilon = 0} \eval{\mathcal{A}_{\text{BFSS}}(\text{out; in}) }_{C_\mu = \varepsilon \partial_\mu \theta_{e,v_s}}. 
\end{equation}
Therefore, the insertion of a single  D0-brane in a momentum eigenstate in the amplitude generates the action of a large $U(1)_{RR}$ gauge transformation \eqref{thta} on the asymptotic scattering state. 
A general gauge transformation can be generated by an appropriate coherent superposition of a momentum-eigenstate D0-brane.

\section*{Acknowledgements}
We would like to thank Alek Bedroya, Alfredo Guevara, Elizabeth Himwich, Patrick Jefferson, Daniel Kapec, Hong Liu, Juan Maldacena, Shu-Heng Shao, and Nicolas Valdes for stimulating discussions. AT and NM gratefully acknowledge support from NSF GRFP grant DGE1745303.

\bibliography{mybib.bib}

\providecommand{\href}[2]{#2}\begingroup\raggedright\begin{thebibliography}{10}

\bibitem{Maldacena:1997re}
J.~M. Maldacena, ``{The Large N limit of superconformal field theories and
  supergravity},'' \href{http://dx.doi.org/10.1023/A:1026654312961}{{\em Adv.
  Theor. Math. Phys.} {\bfseries 2} (1998) 231--252},
  \href{http://arxiv.org/abs/hep-th/9711200}{{\ttfamily arXiv:hep-th/9711200}}.

\bibitem{deWit:1988wri}
B.~de~Wit, J.~Hoppe, and H.~Nicolai, ``{On the Quantum Mechanics of
  Supermembranes},'' \href{http://dx.doi.org/10.1016/0550-3213(88)90116-2}{{\em
  Nucl. Phys. B} {\bfseries 305} (1988) 545}.

\bibitem{Banks_1997}
T.~Banks, W.~Fischler, S.~H. Shenker, and L.~Susskind, ``M theory as a matrix
  model: A conjecture,'' \href{http://dx.doi.org/10.1103/PhysRevD.55.5112}{{\em
  Phys. Rev. D} {\bfseries 55} (Apr, 1997) 5112--5128}.
  \url{https://link.aps.org/doi/10.1103/PhysRevD.55.5112}.

\bibitem{Susskind:1997cw}
L.~Susskind, ``{Another conjecture about M(atrix) theory},''
  \href{http://arxiv.org/abs/hep-th/9704080}{{\ttfamily arXiv:hep-th/9704080}}.

\bibitem{Seiberg_1997}
N.~Seiberg, ``{Why is the matrix model correct?},''
  \href{http://dx.doi.org/10.1103/PhysRevLett.79.3577}{{\em Phys. Rev. Lett.}
  {\bfseries 79} (1997) 3577--3580},
  \href{http://arxiv.org/abs/hep-th/9710009}{{\ttfamily arXiv:hep-th/9710009}}.

\bibitem{Sen:1997we}
A.~Sen, ``{D0-branes on T**n and matrix theory},''
  \href{http://dx.doi.org/10.4310/ATMP.1998.v2.n1.a2}{{\em Adv. Theor. Math.
  Phys.} {\bfseries 2} (1998) 51--59},
  \href{http://arxiv.org/abs/hep-th/9709220}{{\ttfamily arXiv:hep-th/9709220}}.

\bibitem{Polchinski:1999br}
J.~Polchinski, ``{M theory and the light cone},''
  \href{http://dx.doi.org/10.1143/PTPS.134.158}{{\em Prog. Theor. Phys. Suppl.}
  {\bfseries 134} (1999) 158--170},
  \href{http://arxiv.org/abs/hep-th/9903165}{{\ttfamily arXiv:hep-th/9903165}}.

\bibitem{Taylor_2001}
W.~Taylor, ``{M(atrix) Theory: Matrix Quantum Mechanics as a Fundamental
  Theory},'' \href{http://dx.doi.org/10.1103/RevModPhys.73.419}{{\em Rev. Mod.
  Phys.} {\bfseries 73} (2001) 419--462},
  \href{http://arxiv.org/abs/hep-th/0101126}{{\ttfamily arXiv:hep-th/0101126}}.

\bibitem{bigatti1999review}
D.~Bigatti and L.~Susskind, ``{Review of matrix theory},'' {\em NATO Sci. Ser.
  C} {\bfseries 520} (1999) 277--318,
  \href{http://arxiv.org/abs/hep-th/9712072}{{\ttfamily arXiv:hep-th/9712072}}.

\bibitem{Ydri:2017ncg}
B.~Ydri, ``{Review of M(atrix)-Theory, Type IIB Matrix Model and Matrix String
  Theory},'' \href{http://arxiv.org/abs/1708.00734}{{\ttfamily arXiv:1708.00734
  [hep-th]}}.

\bibitem{deBoer:2003vf}
J.~de~Boer and S.~N. Solodukhin, ``{A Holographic reduction of Minkowski
  space-time},'' \href{http://dx.doi.org/10.1016/S0550-3213(03)00494-2}{{\em
  Nucl. Phys. B} {\bfseries 665} (2003) 545--593},
  \href{http://arxiv.org/abs/hep-th/0303006}{{\ttfamily arXiv:hep-th/0303006}}.

\bibitem{He:2015zea}
T.~He, P.~Mitra, and A.~Strominger, ``{2D Kac-Moody Symmetry of 4D Yang-Mills
  Theory},'' \href{http://dx.doi.org/10.1007/JHEP10(2016)137}{{\em JHEP}
  {\bfseries 10} (2016) 137}, \href{http://arxiv.org/abs/1503.02663}{{\ttfamily
  arXiv:1503.02663 [hep-th]}}.

\bibitem{Pasterski:2016qvg}
S.~Pasterski, S.-H. Shao, and A.~Strominger, ``{Flat Space Amplitudes and
  Conformal Symmetry of the Celestial Sphere},''
  \href{http://dx.doi.org/10.1103/PhysRevD.96.065026}{{\em Phys. Rev. D}
  {\bfseries 96} no.~6, (2017) 065026},
  \href{http://arxiv.org/abs/1701.00049}{{\ttfamily arXiv:1701.00049
  [hep-th]}}.

\bibitem{Strominger:2017zoo}
A.~Strominger, ``{Lectures on the Infrared Structure of Gravity and Gauge
  Theory},'' \href{http://arxiv.org/abs/1703.05448}{{\ttfamily arXiv:1703.05448
  [hep-th]}}.

\bibitem{Raclariu:2021zjz}
A.-M. Raclariu, ``{Lectures on Celestial Holography},''
  \href{http://arxiv.org/abs/2107.02075}{{\ttfamily arXiv:2107.02075
  [hep-th]}}.

\bibitem{Pasterski_2021}
S.~Pasterski, ``{Lectures on celestial amplitudes},''
  \href{http://dx.doi.org/10.1140/epjc/s10052-021-09846-7}{{\em Eur. Phys. J.
  C} {\bfseries 81} no.~12, (2021) 1062},
  \href{http://arxiv.org/abs/2108.04801}{{\ttfamily arXiv:2108.04801
  [hep-th]}}.

\bibitem{Taylor:1999gq}
W.~Taylor and M.~Van~Raamsdonk, ``{Multiple D0-branes in weakly curved
  backgrounds},'' \href{http://dx.doi.org/10.1016/S0550-3213(99)00431-9}{{\em
  Nucl. Phys. B} {\bfseries 558} (1999) 63--95},
  \href{http://arxiv.org/abs/hep-th/9904095}{{\ttfamily arXiv:hep-th/9904095}}.

\bibitem{Strominger:2013lka}
A.~Strominger, ``{Asymptotic Symmetries of Yang-Mills Theory},''
  \href{http://dx.doi.org/10.1007/JHEP07(2014)151}{{\em JHEP} {\bfseries 07}
  (2014) 151}, \href{http://arxiv.org/abs/1308.0589}{{\ttfamily arXiv:1308.0589
  [hep-th]}}.

\bibitem{Marotta:2019cip}
R.~Marotta and M.~Verma, ``{Soft Theorems from Compactification},''
  \href{http://dx.doi.org/10.1007/JHEP02(2020)008}{{\em JHEP} {\bfseries 02}
  (2020) 008}, \href{http://arxiv.org/abs/1911.05099}{{\ttfamily
  arXiv:1911.05099 [hep-th]}}.

\bibitem{Ferko:2021bym}
C.~Ferko, G.~Satishchandran, and S.~Sethi, ``{Gravitational memory and compact
  extra dimensions},''
  \href{http://dx.doi.org/10.1103/PhysRevD.105.024072}{{\em Phys. Rev. D}
  {\bfseries 105} no.~2, (2022) 024072},
  \href{http://arxiv.org/abs/2109.11599}{{\ttfamily arXiv:2109.11599 [gr-qc]}}.

\bibitem{Townsend:1995kk}
P.~K. Townsend, ``{The eleven-dimensional supermembrane revisited},''
  \href{http://dx.doi.org/10.1016/0370-2693(95)00397-4}{{\em Phys. Lett. B}
  {\bfseries 350} (1995) 184--187},
  \href{http://arxiv.org/abs/hep-th/9501068}{{\ttfamily arXiv:hep-th/9501068}}.

\bibitem{Witten:1995ex}
E.~Witten, ``{String theory dynamics in various dimensions},''
  \href{http://dx.doi.org/10.1016/0550-3213(95)00158-O}{{\em Nucl. Phys. B}
  {\bfseries 443} (1995) 85--126},
  \href{http://arxiv.org/abs/hep-th/9503124}{{\ttfamily arXiv:hep-th/9503124}}.

\bibitem{Danielsson:1996uw}
U.~H. Danielsson, G.~Ferretti, and B.~Sundborg, ``{D particle dynamics and
  bound states},'' \href{http://dx.doi.org/10.1142/S0217751X96002492}{{\em Int.
  J. Mod. Phys. A} {\bfseries 11} (1996) 5463--5478},
  \href{http://arxiv.org/abs/hep-th/9603081}{{\ttfamily arXiv:hep-th/9603081}}.

\bibitem{Kabat:1996cu}
D.~N. Kabat and P.~Pouliot, ``{A Comment on zero-brane quantum mechanics},''
  \href{http://dx.doi.org/10.1103/PhysRevLett.77.1004}{{\em Phys. Rev. Lett.}
  {\bfseries 77} (1996) 1004--1007},
  \href{http://arxiv.org/abs/hep-th/9603127}{{\ttfamily arXiv:hep-th/9603127}}.

\bibitem{Bachas:1995kx}
C.~Bachas, ``{D-brane dynamics},''
  \href{http://dx.doi.org/10.1016/0370-2693(96)00238-9}{{\em Phys. Lett. B}
  {\bfseries 374} (1996) 37--42},
  \href{http://arxiv.org/abs/hep-th/9511043}{{\ttfamily arXiv:hep-th/9511043}}.

\bibitem{Sethi_1998}
S.~Sethi and M.~Stern, ``{D-brane bound states redux},''
  \href{http://dx.doi.org/10.1007/s002200050374}{{\em Commun. Math. Phys.}
  {\bfseries 194} (1998) 675--705},
  \href{http://arxiv.org/abs/hep-th/9705046}{{\ttfamily arXiv:hep-th/9705046}}.

\bibitem{moore2000d}
G.~Moore, N.~Nekrasov, and S.~Shatashvili, ``D-particle bound states and
  generalized instantons,'' {\em Communications in Mathematical Physics}
  {\bfseries 209} no.~1, (2000) 77--95.

\bibitem{yi1997witten}
P.~Yi, ``Witten index and threshold bound states of d-branes,'' {\em Nuclear
  Physics B} {\bfseries 505} no.~1-2, (1997) 307--318.

\bibitem{plefka1998quantum}
J.~Plefka and A.~Waldron, ``{On the quantum mechanics of M(atrix) theory},''
  \href{http://dx.doi.org/10.1016/S0550-3213(97)00762-1}{{\em Nucl. Phys. B}
  {\bfseries 512} (1998) 460--484},
  \href{http://arxiv.org/abs/hep-th/9710104}{{\ttfamily arXiv:hep-th/9710104}}.

\bibitem{Plefka_1998}
J.~C. Plefka, M.~Serone, and A.~K. Waldron, ``{The Matrix theory S matrix},''
  \href{http://dx.doi.org/10.1103/PhysRevLett.81.2866}{{\em Phys. Rev. Lett.}
  {\bfseries 81} (1998) 2866--2869},
  \href{http://arxiv.org/abs/hep-th/9806081}{{\ttfamily arXiv:hep-th/9806081}}.

\bibitem{Plefka:1998in}
J.~Plefka, M.~Serone, and A.~Waldron, ``{Matrix theory and Feynman diagrams},''
  \href{http://dx.doi.org/10.1002/(SICI)1521-3978(20001)48:1/3<191::AID-PROP191>3.0.CO;2-#}{{\em
  Fortsch. Phys.} {\bfseries 48} (2000) 191--194},
  \href{http://arxiv.org/abs/hep-th/9903099}{{\ttfamily arXiv:hep-th/9903099}}.

\bibitem{Plefka:1997hm}
J.~Plefka and A.~Waldron, ``{Asymptotic supergraviton states in matrix
  theory},'' in {\em {31st International Ahrenshoop Symposium on the Theory of
  Elementary Particles}}, pp.~130--136.
\newblock 9, 1997.
\newblock \href{http://arxiv.org/abs/hep-th/9801093}{{\ttfamily
  arXiv:hep-th/9801093}}.

\bibitem{Becker:2006dvp}
K.~Becker, M.~Becker, and J.~H. Schwarz,
  \href{http://dx.doi.org/10.1017/CBO9780511816086}{{\em {String theory and
  M-theory: A modern introduction}}}.
\newblock Cambridge University Press, 12, 2006.

\bibitem{Becker:1997wh}
K.~Becker and M.~Becker, ``{A Two loop test of M(atrix) theory},''
  \href{http://dx.doi.org/10.1016/S0550-3213(97)00518-X}{{\em Nucl. Phys. B}
  {\bfseries 506} (1997) 48--60},
  \href{http://arxiv.org/abs/hep-th/9705091}{{\ttfamily arXiv:hep-th/9705091}}.

\bibitem{Becker:1997xw}
K.~Becker, M.~Becker, J.~Polchinski, and A.~A. Tseytlin, ``{Higher order
  graviton scattering in M(atrix) theory},''
  \href{http://dx.doi.org/10.1103/PhysRevD.56.R3174}{{\em Phys. Rev. D}
  {\bfseries 56} (1997) R3174--R3178},
  \href{http://arxiv.org/abs/hep-th/9706072}{{\ttfamily arXiv:hep-th/9706072}}.

\bibitem{Weinberg:1965nx}
S.~Weinberg, ``{Infrared photons and gravitons},''
  \href{http://dx.doi.org/10.1103/PhysRev.140.B516}{{\em Phys. Rev.} {\bfseries
  140} (1965) B516--B524}.

\bibitem{Kapec_2017}
D.~Kapec, V.~Lysov, S.~Pasterski, and A.~Strominger, ``{Higher-dimensional
  supertranslations and Weinberg\textquoteright{}s soft graviton theorem},''
  \href{http://dx.doi.org/10.4310/AMSA.2017.v2.n1.a2}{{\em Ann. Math. Sci.
  Appl.} {\bfseries 02} (2017) 69--94},
  \href{http://arxiv.org/abs/1502.07644}{{\ttfamily arXiv:1502.07644 [gr-qc]}}.

\bibitem{He_2019_1}
T.~He and P.~Mitra, ``{Asymptotic symmetries and Weinberg\textquoteright{}s
  soft photon theorem in Mink$_{d+2}$},''
  \href{http://dx.doi.org/10.1007/JHEP10(2019)213}{{\em JHEP} {\bfseries 10}
  (2019) 213}, \href{http://arxiv.org/abs/1903.02608}{{\ttfamily
  arXiv:1903.02608 [hep-th]}}.

\bibitem{He_2019}
T.~He and P.~Mitra, ``{Asymptotic symmetries in (d + 2)-dimensional gauge
  theories},'' \href{http://dx.doi.org/10.1007/JHEP10(2019)277}{{\em JHEP}
  {\bfseries 10} (2019) 277}, \href{http://arxiv.org/abs/1903.03607}{{\ttfamily
  arXiv:1903.03607 [hep-th]}}.

\bibitem{Kapec_2022}
D.~Kapec and P.~Mitra, ``{Shadows and soft exchange in celestial CFT},''
  \href{http://dx.doi.org/10.1103/PhysRevD.105.026009}{{\em Phys. Rev. D}
  {\bfseries 105} no.~2, (2022) 026009},
  \href{http://arxiv.org/abs/2109.00073}{{\ttfamily arXiv:2109.00073
  [hep-th]}}.

\bibitem{He:2014cra}
T.~He, P.~Mitra, A.~P. Porfyriadis, and A.~Strominger, ``{New Symmetries of
  Massless QED},'' \href{http://dx.doi.org/10.1007/JHEP10(2014)112}{{\em JHEP}
  {\bfseries 10} (2014) 112}, \href{http://arxiv.org/abs/1407.3789}{{\ttfamily
  arXiv:1407.3789 [hep-th]}}.

\bibitem{Campiglia:2015lxa}
M.~Campiglia, ``{Null to time-like infinity Green\textquoteright{}s functions
  for asymptotic symmetries in Minkowski spacetime},''
  \href{http://dx.doi.org/10.1007/JHEP11(2015)160}{{\em JHEP} {\bfseries 11}
  (2015) 160}, \href{http://arxiv.org/abs/1509.01408}{{\ttfamily
  arXiv:1509.01408 [hep-th]}}.

\bibitem{Kapec:2015ena}
D.~Kapec, M.~Pate, and A.~Strominger, ``{New Symmetries of QED},''
  \href{http://dx.doi.org/10.4310/ATMP.2017.v21.n7.a7}{{\em Adv. Theor. Math.
  Phys.} {\bfseries 21} (2017) 1769--1785},
  \href{http://arxiv.org/abs/1506.02906}{{\ttfamily arXiv:1506.02906
  [hep-th]}}.

\bibitem{Campiglia:2015qka}
M.~Campiglia and A.~Laddha, ``{Asymptotic symmetries of QED and
  Weinberg\textquoteright{}s soft photon theorem},''
  \href{http://dx.doi.org/10.1007/JHEP07(2015)115}{{\em JHEP} {\bfseries 07}
  (2015) 115}, \href{http://arxiv.org/abs/1505.05346}{{\ttfamily
  arXiv:1505.05346 [hep-th]}}.

\bibitem{douglas1997d}
M.~R. Douglas, D.~Kabat, P.~Pouliot, and S.~H. Shenker, ``D-branes and short
  distances in string theory,'' {\em Nuclear Physics B} {\bfseries 485}
  no.~1-2, (1997) 85--127.

\end{thebibliography}\endgroup
\bibliographystyle{utphys}


\end{document}